\documentclass[aps,prd,amsmath,twocolumn,groupedaddress,showkeys,showpacs]{revtex4-1}
\usepackage{graphicx}
\usepackage{epsfig}
\newcommand{\beqy}{\begin{eqnarray}}
\newcommand{\eeqy}{\end{eqnarray}}

\begin{document}

\title{Stability of super-Chandrasekhar magnetic white dwarfs}
\author{N. Chamel}
\author{A.~F. Fantina}
\author{P.~J. Davis}
\affiliation{Institut d'Astronomie et d'Astrophysique, CP-226, Universit\'e Libre de Bruxelles, 
1050 Brussels, Belgium}

\begin{abstract}
It has been recently proposed that very massive white dwarfs endowed with strongly quantizing magnetic fields
might be the progenitors of overluminous type Ia supernovae like SN~2006gz and SN~2009dc. In this work, we show 
that the onset of electron captures and pycnonuclear reactions in these putative super-Chandrasekhar white 
dwarfs may severely limit their stability. 
\end{abstract}



\maketitle

{\it Introduction.} With a mass comparable to that of our Sun but enclosed inside a radius of the order of that 
of Earth, white dwarfs represent the final stage of low and intermediate mass stars (i.e., with a mass 
$\lesssim 10 M_\odot$, $M_\odot$ being the mass of our Sun). As a consequence, their observations 
can potentially shed light on stellar, galactic and even cosmic evolution (see, e.g., Ref.~\cite{koest1990}). 
In particular, white dwarfs are 
generally thought to be the progenitors of type Ia supernova (SNIa) explosions (see, e.g., Ref.~\cite{hill2013} 
and references therein for a recent review). With their assumed universal calibrated light-curves, SNIa have been widely 
used as distance indicators in observational cosmology thus showing that the expansion of the Universe is accelerating
(see, e.g., Ref.~\cite{perl2012}). However, the recent discovery of overluminous SNIa, whose progenitors are thought to be 
``super-Chandrasekhar'' white dwarfs with a mass $M>2M_\odot$, may spoil cosmological 
measurements (see, e.g., Ref.~\cite{hill2013}). These peculiar SNIa may result from the explosions of rapidly rotating white dwarfs 
or from the mergers of two massive white dwarfs~\cite{how2006}. Alternatively, it has been recently suggested that overluminous 
SNIa could be the manifestations of the explosions of strongly magnetized white dwarfs~\cite{km2012,dm2012,dm2012b,dm2013,dm2013b}. 
In the presence of a strongly quantizing magnetic field, the equation of state of dense matter becomes so stiff that the 
electron degeneracy pressure may be high enough to support white dwarfs with masses up to $\sim 2.6 M_\odot$. In this paper, 
we show that the onset of electron captures and pycnonuclear reactions, which were neglected in previous studies, may severely limit 
the stability of such putative super-Chandrasekhar white dwarfs. The global stability of such stars is also briefly discussed. 

{\it Model of magnetic white-dwarf matter.} The global structure of a white dwarf is determined by the equation of state, i.e. 
the relation between the pressure $P$ and the mass density $\rho$. In the model we adopt here (see, e.g., Ref.~\cite{chapav12} and 
references therein), the interior of the star is assumed to be made of fully ionized atoms arranged in a regular crystal lattice 
at zero temperature. In addition, the whole star is supposed to contain crystalline structures made of the same type of nuclides 
with proton number $Z$ and atomic number $A$ (but the geometry of the crystal lattice can vary with depth). The electric charge 
neutrality condition implies that mass density can be expressed as $\rho=n_e m A/Z $ where $n_e$ is the electron number density 
and $m$ the average mass per nucleon (approximated here by the unified atomic mass unit). 
The pressure $P$ 
is given by the sum of the electron degeneracy pressure $P_e$ and the pressure $P_L$ arising from the crystal lattice. According to 
the Bohr-van Leeuwen theorem~\cite{bvl32}, the lattice pressure is independent of the magnetic field apart from a negligibly small 
contribution due to quantum zero-point motion of ions~\cite{bai09}. In the high-density domain that we consider here $P_L\ll P_e$ 
(see, e.g., Ref.~\cite{chapav12}). In the following, we will therefore approximate $P$ by $P_e$. In the presence of a strong magnetic 
field, the electron motion perpendicular to the field is quantized into Landau levels (see, e.g., chapter 4 from Ref.~\cite{hae07}). Ignoring the electron anomalous magnetic moment (see, e.g., Section 4.1.1 from Ref.~\cite{hae07} and references therein), and treating electrons as a relativistic Fermi gas, the energies of 
Landau levels (which were actually first found by Rabi~\cite{rab28}) are given by
\begin{equation}
e_{\nu} = \sqrt{c^2 p_z^2+m_e^2 c^4(1+2\nu B_\star)}
\end{equation}
\begin{equation}
\nu = n_L + \frac{1}{2}+\sigma\, ,
\end{equation}
where $m_e$ is the electron mass, $c$ is the speed of light, $n_L$ is any non-negative integer, $\sigma=\pm 1/2$ is the spin, $p_z$ is 
the component of the momentum along the field, and $B_\star=B/B_\textrm{crit}$ with the critical magnetic field $B_\textrm{crit}$ defined by 
$B_\textrm{crit}=m_e^2 c^3/(e\hbar)\simeq 4.4\times 10^{13}$~G, $e$ being the elementary charge (gaussian cgs units are used throughout this paper). 
The expressions for the electron energy density $\mathcal{E}_e$ and corresponding 
electron pressure $P_e$ can be found in Ref.~\cite{hae07}. 

In the outer shell of a real white dwarf, matter is generally not degenerate and is expected to contain various light elements, 
mainly hydrogen and helium. However, the contribution of this region to the maximum mass, which is our main concern here, is 
negligible (see, e.g., Ref.~\cite{wil89}). 

{\it Maximum mass of super-Chandrasekhar white dwarfs.}
Previous studies showed that the maximum mass of magnetic white dwarfs is almost unchanged for $B_\star\leq 1$~\cite{wil89,suh00,bed03}. 
While in the absence of a magnetic field $P\propto\rho^{4/3}$ at high densities, the equation of state becomes much stiffer in the 
strongly quantizing regime $B_\star\gg 1$, with the pressure varying as $P\propto\rho^2$. For this reason, it has been recently argued that the 
maximum mass of strongly magnetized white dwarfs could be much higher than the Chandrasekhar limit for ordinary white dwarfs~\cite{km2012,dm2012,dm2012b,dm2013,dm2013b}. 
Considering a uniform magnetic field, these authors showed that the limiting mass of such hypothetical super-Chandrasekhar magnetic white dwarfs is reached when the 
central density $\rho_c$ approximately coincides with the density $\rho_{\rm B}$ at which electrons start to populate the level 
$\nu=1$. This density is given by (see, e.g., Chap. 4 in Ref.~\cite{hae07})
\begin{equation}\label{eq:rhoB}
\rho_{\rm B}(A,Z,B_\star) =\frac{A}{Z} \frac{m}{\lambda_e^3} \frac{B_\star^{3/2}}{\sqrt{2} \pi^2} \, ,
\end{equation} 
where $\lambda_e=\hbar/(m_e c)$ is the electron Compton wavelength. 
The authors of Ref.~\cite{dm2013} estimated the maximum mass of a white dwarf endowed with a uniform magnetic field from the analytic solutions 
of the Lane-Emden equation. Considering the strongly quantizing regime for which $B_\star\rightarrow +\infty$ so $P\approx K_m \rho^2$, with 
$K_m=Z^2 m_e c^2 \pi^2 \lambda_e^3/(A^2 m^2 B_\star)$, they found the following expressions for the white dwarf mass 
$M_0$ and radius $R_0$
\begin{equation}\label{eq:lane-emden}
M_0=\sqrt{2\pi}\rho_c \left(\frac{K_m}{G}\right)^{3/2}\, ,\hskip0.5cm R_0=\sqrt{\frac{\pi K_m}{2G}}\, ,
\end{equation}
where $G$ is the gravitational constant. Note that the radius is independent of $\rho_c$, and is completely determined by 
the stellar composition and the magnetic field strength. Setting $\rho_c=\rho_{\rm B}$ in Eq.~(\ref{eq:lane-emden}) using 
Eq.~(\ref{eq:rhoB}) leads to the maximum mass estimate~\cite{dm2013} 
\begin{equation}\label{eq:Mmax}
M_{\rm max}(A,Z) = \left(\frac{Z}{A}\right)^2\left(\frac{\pi \hbar c}{G}\right)^{3/2} \frac{1}{m^2}\, .
\end{equation}
Note that $M_{\rm max}$ is independent of the magnetic field strength. Replacing $Z/A=1/2$ in Eq.~(\ref{eq:Mmax}) for a 
carbon-oxygen white dwarf yields $M_{\rm max}\simeq 2.6 M_\odot$. However, as recently discussed in Ref.~\cite{nit2013,coelho13}, such stars would undergo 
global instabilities due to the huge magnetic forces exerted on their surface.

{\it Global stability of super-Chandrasekhar white dwarfs.} For a stellar configuration to be stable, 
Chandrasekhar and Fermi~\cite{chandra53} showed a long time ago that the magnetic energy
\begin{equation}\label{eq:Emag}
E_{\rm mag}=\frac{1}{8\pi}\int_0^M \frac{B^2}{\rho} dm \, ,
\end{equation}
should be lower than the absolute value of the gravitational energy 
\begin{equation}\label{eq:Egrav}
E_{\rm grav}=\frac{1}{2} \int_0^M \Phi dm \, ,
\end{equation}
where $\Phi$ is the gravitational potential. For the super-Chandrasekhar white dwarf configurations obtained in Ref.~\cite{dm2013}, 
using the analytic solutions of the Lane-Emden equation we find that 
\begin{equation}\label{eq:global-stability}
\frac{E_{\rm mag}}{|E_{\rm grav}|}=\frac{\pi^3}{18\alpha}\simeq 236 \, ,
\end{equation}
where $\alpha = e^2/\hbar c$ is the fine-structure constant. Therefore, we must conclude that a star endowed with a \emph{uniform} and strongly 
quantizing magnetic field is globally unstable. Note that this conclusion is independent of the magnetic field strength, as shown by Eq.~(\ref{eq:global-stability}). 
However, the existence of super-Chandrasekhar magnetic white dwarfs is not necessarily ruled out. Indeed, one may assume that the magnetic field is the strongest 
at the center of the star, and decreases outwards in such a way that the global stability condition $E_{\rm mag}<|E_{\rm grav}|$ is fulfilled. On the other hand, 
such stars could still be \emph{locally} unstable due to matter neutronization. 

{\it Matter neutronization.}
With increasing pressure, a nucleus $X$ with proton number $Z$ and atomic number $A$ may become unstable 
and transform into a nucleus $Y$ with proton number $Z-1$ and atomic number $A$ through the capture of an electron
and the emission of a neutrino~:
\begin{equation}\label{eq:e-capture}
^A_ZX+e^- \rightarrow ^A_{Z-1}Y+\nu_e\, .
\end{equation}
The nucleus $Y$ itself is generally unstable and captures another electron. Electron captures lead to a strong softening of 
the equation of state, since the pressure remains unchanged while the mass density increases from $\rho=n_e m A/Z $ to 
$\rho=n_e m A/(Z-2)$ (note that $n_e$ is essentially constant during the transition since $P\approx P_e(n_e)$). It is well-known
that such a softening limits the stability of ordinary white dwarfs, as first pointed out by Gamow~\cite{gamow39} and Chandrasekhar~\cite{chandra41} a long time ago  (see also  Refs.~\cite{sch58,htww65}). 

For the process~(\ref{eq:e-capture}) to occur in a layer of the star characterized by a pressure $P$, 
the following inequality has to be fulfilled:
\begin{equation}\label{eq:e-capture-gibbs}
g(P,A,Z)\geq g(P,A,Z-1)\, ,
\end{equation}
where $g(P,A,Z)$ denotes the Gibbs free energy per nucleon for nuclei with proton number $Z$ and mass number $A$, 
given by (see, e.g., Ref.~\cite{chapav12} and references therein)
\begin{equation}\label{eq:gibbs}
g(P,A,Z)=m c^2+\frac{\Delta(A,Z)}{A}+\frac{Z}{A}\left(\mu_e - m_e c^2 \right)
\end{equation}
$\Delta(A,Z)$ being the mass excess of the nucleus, and 
$\mu_e=d\mathcal{E}_e/dn_e$ is the electron chemical potential. The lattice contribution to $g$ is small and has thus been ignored. It 
mostly affects the equation of state at low densities (see, e.g., Ref.~\cite{chapav12}). 
In principle, $\Delta(A,Z)$ could also depend on the magnetic field. However, the effect induced by the magnetic field is 
very small for $B\lesssim 10^{17}$~G~\cite{pen11}. 

Note that for an isolated nucleus $\mu_e=m_e c^2$ so Eq.~(\ref{eq:e-capture-gibbs}) reduces to the familiar condition 
$\Delta(A,Z)\geq \Delta(A,Z-1)$. 
Even though a nucleus in a vacuum could be stable, it will generally undergo electron capture in matter at high enough densities. Combining 
Eqs.~(\ref{eq:e-capture-gibbs}) and (\ref{eq:gibbs}), we find the following condition for electron capture to occur: 
\begin{equation}\label{eq:muebeta}
\mu_e \geq \mu_e^\beta(A,Z)\equiv \Delta(A,Z-1)-\Delta(A,Z) + m_e c^2\, .
\end{equation}
In the absence of magnetic fields and in the limit of ultrarelativistic electrons, $\mu_e\approx \hbar c (3 \pi^2 n_e)^{1/3}$ 
so the threshold mass density for matter to become unstable against electron captures is approximately given by 
\begin{equation}\label{eq:rhobeta0}
\rho_\beta^0(A,Z) \approx \frac{A}{Z} \frac{m}{3\pi^2 (\hbar c)^3} \mu_e^\beta(A,Z)^3\, .
\end{equation}
In the presence of a strongly quantizing magnetic field such that only the lowest level $\nu=0$ is filled, 
$\mu_e\approx 2 \pi^2 m_e c^2 \lambda_e^3 n_e/B_\star$ (see, e.g., Ref.~\cite{chapav12}) and the threshold mass density thus becomes
\begin{equation}\label{eq:rhobeta}
\rho_\beta(A,Z,B_\star) \approx \frac{A}{Z} \frac{m B_\star}{2\pi^2 \lambda_e^3 } \frac{ \mu_e^\beta(A,Z)}{m_e c^2}\, .
\end{equation}

{\it Local stability of super-Chandrasekhar white dwarfs.} The central density of such stars will be generally limited by 
the softening of the equation of state due to the complete filling of the lowest Landau level. However, if 
$\rho_\beta(A,Z,B_\star) < \rho_{\rm B}(A,Z,B_\star)$ at the center of the star, or equivalently 
\begin{equation}\label{eq:Bstarmax}
B_\star >B_\star^\beta(A,Z) \equiv \frac{1}{2} \left(\frac{ \mu_e^\beta(A,Z)}{m_e c^2}\right)^2 
\end{equation}
after combining Eqs.~(\ref{eq:rhobeta}) and (\ref{eq:rhoB}), the stellar matter will become locally 
unstable against electron capture. This instability will be accompanied by a strong softening of the equation 
of state, thus limiting the central density of the star. 

The maximum maximorum of the white dwarf mass will therefore be 
reached for $B_\star \approx B_\star^\beta$. 
Note that $\mu_e^\beta$, and hence also $B_\star^\beta$, are uniquely determined by nuclear masses. Most white dwarfs are 
expected to have cores composed mainly of carbon and oxygen, the primary ashes of helium burning. However, some white dwarfs 
may contain other elements in their cores like helium~\cite{nel98,lieb04,ben05}, neon and magnesium~\cite{nom84} 
or even much heavier elements like iron~\cite{pan00}. The masses of these nuclei have all been measured in the laboratory. 
Using the latest experimental data from the Atomic Mass Evaluation 2012~\cite{ame2012}, we thus find $B_\star^\beta\simeq 853$ 
for $^{4}$He~\cite{helium}, 369 for $^{12}$C, 229 for $^{16}$O, 109 for $^{20}$Ne, 70 for $^{24}$Mg, 
and 34 for $^{56}$Fe (the nuclear masses can be obtained from the tabulated \emph{atomic} masses after subtracting out the 
binding energy of the atomic electrons; however the differences are very small~\cite{lpt03} and have been ignored here). 

We have thus shown that the putative super-Chandrasekhar magnetic white dwarfs considered in Refs.~\cite{km2012,dm2012,dm2012b,dm2013,dm2013b} 
are only stable against electron capture provided $B_\star < B_\star^\beta$. On the other hand, the star 
could still remain in a \emph{metastable} state for $B_\star > B_\star^\beta$ since the magnetic field can strongly reduce 
the electron capture rate~\cite{lai91}. However, this reduction is most effective at very low densities $\rho\ll \rho_{\rm B}$~\cite{lai91}; 
in the stellar core at densities $\rho\approx\rho_{\rm B}$ nuclei will therefore capture electrons at essentially the same 
rate as in ordinary white dwarfs. Let us consider the carbon-oxygen white-dwarf configuration discussed in Ref.~\cite{dm2013} 
with the mass $M\simeq 2.6 M_\odot$ and magnetic field $B_\star\simeq 2\times 10^4$. Using the nuclear model described in 
Refs.~\cite{paar09,fant12}, we find that carbon and oxygen would capture electrons at a rate of order $10^{4}$  
events per second at the center of the star where $\rho_c \approx \rho_{\rm B}$. Therefore, such a star would be highly unstable. 
In addition, the star would also be unstable against 
neutron emission. Indeed, this process becomes energetically favorable when the Gibbs free energy per nucleon $g$ exceeds the 
neutron rest mass energy. Using Eq.~(\ref{eq:gibbs}) with $\mu_e\approx 2\pi^2 m_e c^2 \lambda_e^3/B_\star$ (strongly quantizing magnetic
field), we find that neutron emission can occur in the stellar core if $B_\star > 531$ for carbon and $B_\star > 570$ for oxygen, 
magnetic field strengths much lower than the value $B_\star\simeq 2\times 10^4$ assumed to prevail in the star.

{\it Pycnonuclear reactions}. At sufficiently high densities, the quantum-zero point fluctuations of nuclei about their 
equilibrium position may become large enough to trigger pycnonuclear fusion reactions
\begin{equation}\label{eq:pycno}
 ^A_ZX + ^A_ZX \rightarrow ^{2A}_{2Z}Y\, .
\end{equation}
However, the rates at which these processes occur still remain very uncertain (see, e.g., Ref.~\cite{yak06}). Let us simply consider 
that nuclei will fuse at densities $\rho>\rho_{\rm pyc}$. For a given magnetic field, the threshold density for the onset of 
electron capture is generally lower for the daughter nucleus $^{2A}_{2Z}Y$ than for the original nucleus $^A_ZX$. For this reason, 
pycnonuclear reactions further limit the stability of white dwarfs whenever $\rho_{\rm pyc}<\rho_\beta(2A,2Z,B_\star)$, assuming 
$\rho_\beta(2A,2Z,B_\star)<\rho_c$. As previously discussed, 
the maximum maximorum of the white-dwarf mass will be reached when electrons completely fill the lowest level $\nu=0$, i.e. for magnetic 
field strength $B_\star\approx B_\star^\beta(2A,2Z)$ such that $\rho_c\approx \rho_{\rm B}(2A,2Z,B_\star)\approx\rho_\beta(2A,2Z,B_\star)$. Examples of values of 
$B_\star^\beta(2A,2Z)$ are 70 for $^{24}$Mg (from the fusion of $^{12}$C), 9.5 for $^{32}$S (from the fusion of $^{16}$O), and 6.4 
for $^{40}$Ca (from the fusion of $^{20}$Ne).
Thus it can be seen that the largest possible magnetic field strength of hypothetical super-Chandrasekhar white dwarfs could be even 
lower than those discussed in the previous section if pycnonuclear reactions are allowed. These magnetic field strengths are also 
considerably lower than the range of values $B_\star\sim 10^2-10^4$ assumed in previous studies of super-Chandrasekhar 
magnetic white dwarfs~\cite{km2012,dm2012,dm2012b,dm2013,dm2013b} thus casting doubts on the existence of such stars.
On the other hand, substituting $B_\star^\beta(2A,2Z)$ in Eq.~(\ref{eq:rhoB}), we find that pycnonuclear fusions will not further 
reduce the stability of white dwarfs if they occur at high enough densities, namely
\begin{equation}\label{eq:rhopycmax}
\rho_{\rm pyc} > \rho_{\rm pyc}^{\rm min}(A,Z)\equiv \frac{B_\star^\beta(2A,2Z)^{3/2}}{\sqrt{2} \pi^2 \lambda_e^3} \frac{A}{Z} m\, .
\end{equation} 
Examples of values of $\rho_{\rm pyc}^{\rm min}$ are $2.4\times 10^9$~g~cm$^{-3}$ for the fusion of $^{12}$C, $1.2\times 10^8$~g~cm$^{-3}$ 
for the fusion of $^{16}$O, and $6.6\times 10^7$~g~cm$^{-3}$ for the fusion of $^{20}$Ne.

{\it Conclusion.} Putative super-Chandrasekhar white dwarfs endowed with \emph{uniform} strongly quantizing magnetic fields, as 
considered in Refs~\cite{km2012,dm2012,dm2012b,dm2013,dm2013b}, are found to be globally unstable (see also Refs.~\cite{nit2013,coelho13}). 
However, this conclusion does not necessarily preclude the existence of super-Chandrasekhar magnetic white dwarfs with spatially varying 
magnetic fields in their interior. On the other hand, the softening of the equation of state accompanying the onset of electron captures and 
pycnonuclear reactions in the core of these stars will lead to \emph{local} instabilities. We have shown that these instabilities set an 
upper limit to the magnetic field strength at the center of the star, ranging from $~10^{14}$~G to $~10^{16}$~G depending on the 
core composition. In particular, pycnonuclear reactions can occur in the core of carbon (oxygen) white dwarfs if the magnetic field strength exceeds 
about $3.1\times 10^{15}$~G ($4.2\times 10^{14}$~G). These values are significantly lower than those considered in Refs~\cite{km2012,dm2012,dm2012b,dm2013,dm2013b}. 
The stability of these stars may be further limited by general relativity~\cite{coelho13} and by the pressure anisotropy induced by the presence of the 
magnetic field~\cite{ferrer2010,dm2012} (but see also Refs.~\cite{pot2012,ferrer2012}). As a matter of fact, observations suggest an upper limit of the surface magnetic field strength 
of $B\approx 10^{9}$~G for isolated white dwarfs~\cite{schmidt03}, while in binary systems, $B$ is typically $1-6\times{10}^{7}$~G~\cite{schwope06}, 
in rare cases exceeding $10^{8}$~G (see, e.g., Ref.~\cite{gaensicke04}). It remains to be shown that stellar configurations with central magnetic fields 
as strong as $10^{15}-10^{16}$~G are stable. Further studies of hydromagnetostatic equilibrium are therefore needed 
before any firm conclusions on the existence of super-Chandrasekhar magnetic white dwarfs can be drawn.

\begin{acknowledgments}
This work was financially supported by FNRS (Belgium). The authors thank S. Goriely for fruitful discussions. 
The authors are also particularly grateful to G. Col\`o, E. Khan, N. Paar and D. Vretenar for providing us with 
their computer codes. 
\end{acknowledgments}


\begin{thebibliography}{99}
\bibitem{helium} In this case, it is energetically favorable to emit free neutrons rather than capturing electrons so that the 
following reaction will actually occur: $^{4}{\rm He}\rightarrow ^{3}{\rm H}+n$.
\end{thebibliography}
\end{document}